\documentclass[preprint]{aastex}
\shorttitle{SZ Herculis: Revised LITE Model and Orbit Stability}
\shortauthors{Hinse et al.}

\usepackage[dvips]{color}
\usepackage{ulem}

\begin{document}

\title {The Proposed Quadruple System SZ Herculis: Revised LITE Model and 
Orbital Stability Study}

\author{Tobias Cornelius Hinse}
\affil{Korea Astronomy and Space Science Institute, Daejeon 305-348, Republic 
of Korea}
\affil{Armagh Observatory, College Hill, BT61 9DG, Armagh, NI, UK}
\email{tchinse@gmail.com}

\author{Krzysztof Go{\'z}dziewski}
\affil{Nicolaus Copernicus University, Torun Centre for Astronomy, PL-87-100 
Torun, Poland}

\author{Jae Woo Lee}
\affil{Korea Astronomy and Space Science Institute, Daejeon 305-348, Republic 
of Korea}

\author{Nader Haghighipour}
\affil{Institute for Astronomy \& NASA Astrobiology Institute, University of 
Hawaii, 96822 HI, USA.}

\author{Chung-Uk Lee}
\affil{Korea Astronomy and Space Science Institute, Daejeon 305-348, Republic 
of Korea}

\begin{abstract}
In a recent study, Lee et al. presented new photometric follow-up timing 
observations of the semi-detached binary system SZ Herculis and proposed the 
existence of two hierarchical cirumbinary companions. Based on the light-travel
time effect, the two low-mass M-dwarf companions are found to orbit the binary 
pair on moderate to high eccentric orbits. The derived periods of these two 
companions are close to a 2:1 mean-motion orbital resonance. We have studied 
the stability of the system using the osculating orbital elements as presented 
by Lee et al. Results indicate an orbit-crossing architecture exhibiting 
short-term dynamical instabilities leading to the escape of one of the proposed
companions. We have examined the system's underlying model parameter-space by 
following a Monte Carlo approach and found an improved fit to the timing data. 
A study of the stability of our best-fitting orbits also indicates that the 
proposed system is generally unstable. If the observed anomalous timing 
variations of the binary period is due to additional circumbinary companions, 
then the resulting system should exhibit a long-term stable orbital 
configuration much different from the orbits suggested by Lee et al. We, 
therefore, suggest that based on Newtonian-dynamical considerations, the 
proposed quadruple system cannot exist. To uncover the true nature of the  
observed period variations of this system, we recommend future photometric 
follow-up observations that could further constrain eclipse-timing variations 
and/or refine light-travel time models.
\end{abstract}

{}

\section{INTRODUCTION}

Formation of multiple star systems is complex and is believed to occur either 
by interaction/capture mechanisms \citep[and references 
therein]{vandenBerkEtAl2007} during the formation and dynamical evolution of 
globular star clusters, or directly from a massive primordial disk involving 
accretion processes or disk instabilities \citep{LimTakakuwa2006, 
DucheneEtAl2007,MarzariEtAl2009}. The exact formation channel is not yet fully 
understood. However, these mechanisms might be capable of producing various 
star systems characterized by different orbital architectures and/or 
hierarchies \citep{Evans1968}.

Cirumbinary objects belong to a category of hierarchical star systems where one
(or more) massive companion(s) orbits a pair of stars. An example of such  
systems is the quadruple (or quaternary) system of HD~98800 consisting of two 
distinct spectroscopic binaries orbiting a common center of mass 
\citep{FurlanEtAl2007}. Single or multiple low-mass circumbinary companions of 
planetary nature have been recently discovered from ground-based observations 
\citep{LeeEtAl2009,BeuermannEtAl2010, PotterEtAl2011, QianEtAl2011} and with 
the {\sc KEPLER} space telescope \citep{DoyleEtAl2011, WelshEtAl2012}. However,
some of these multi-planet circumbinary systems (NN Ser, HW Vir, HU Aqr and DP 
Leo) have proposed orbital properties that seem to render their orbits unstable
\citep{HornerEtAl2011,HinseEtAl2011,WittenmyerEtAl2011,FunkEtAl2011, 
HornerEtAl2012}.

In a recent work, \cite{LeeEtAl2011} presented new photometric observations of 
the Algol-type semi-detached binary star system SZ Herculis (SZ Her(AB) 
\footnote{this notation follows the notation as suggested by 
\cite{HessmanEtAl2010}} hereafter). Based on more than 1,000 eclipse measurements (spanning more than a century) these authors were able to detect a significant change in the systems orbital period manifesting itself as eclipse timing variations (ETVs). Such a change in the binary period can be due to i) the interaction between the two binary components through their magnetic fields, mass-transfer, or tidal interactions (resulting in apsidal motion) ii) gravitational perturbations by additional massive objects (companions) and/or iii) by the light-travel time effect (LITE\footnote{Sometimes referred to as 
LTTE or LTT in the literature}) also known as R{\o}mer effect \citep{Irwin1952, Irwin1959}.

It is important to note that timing measurement errors can be uncorrelated (white noise following a Gaussian distribution) and/or of systematical (correlated or red noise) origin. For timing measurements of pulsars, this has recently been pointed out by \cite{ColesEtAl2011} as a possible cause of 
errors in estimating the model parameters. In the past, neglecting the effect 
of red noise was responsible for the false detection of planets around 
pulsars \citep{BailesEtAl1991}.

The LITE effect implies the presence of one or more massive object(s) which 
can result in the reflex motion of the binary barycenter about the total system's center of mass. This reflex motion (or binary wobble) gives rise to time varying light-travel paths resulting in differences in the periodic mid-eclipse timing. It is important to note that the LITE effect is a geometrical effect and does not involve gravitational perturbations. For instance, in a hierarchical triple stellar system (a third companion orbiting 
a close binary), the third object creates a single-body LITE effect, introducing a sinusoidal-like variation in the binaries orbital period \citep{Irwin1959}.

In their work, \cite{LeeEtAl2011} applied various eclipse timing variation  
models in an attempt to describe the observed period variations of SZ Her. 
Their most promising fit to the times of minimum light points to a two-companion LITE model, implying the existence of two low-mass M-type stars orbiting SZ Her on circumbinary orbits. The orbital periods of these companions are $P_3 \backsimeq 86$ and $P_4 \backsimeq 43$ years, suggesting a near 2:1 mean-motion resonance between their orbits.

In this work, using the fitted orbital parameters from \cite{LeeEtAl2011}, 
we study the stability of the two suggested M-dwarf companions. We then 
perform an extensive parameter-search for a best-fit model using the complete set of timing data of SZ Her as compiled from the literature and transformed 
to the terrestrial time (TT) standard. In particular, we carry out a quasi-global Monte Carlo search of a variety of two-companion LITE models to explore a region of $\chi^2$ parameter-space. Finally, the orbital 
stability of our best-fit model is examined and its implications are 
discussed.

\section{BARYCENTRIC TWO-COMPANION LITE MODEL}

\begin{figure*}
% Tobys internal note:
% Path:/home/tobiash/simulations/SZHer/XFigure
% Xfig file: O_LTTOrbit.fig
% Note: the line \usepackage[dvips]{color} in the preamble of the LaTeX file 
% is necessary to make this (*.pstex_t) work. Also within XFig use "special" 
% flag.
\centering
\scalebox{0.70}{\input{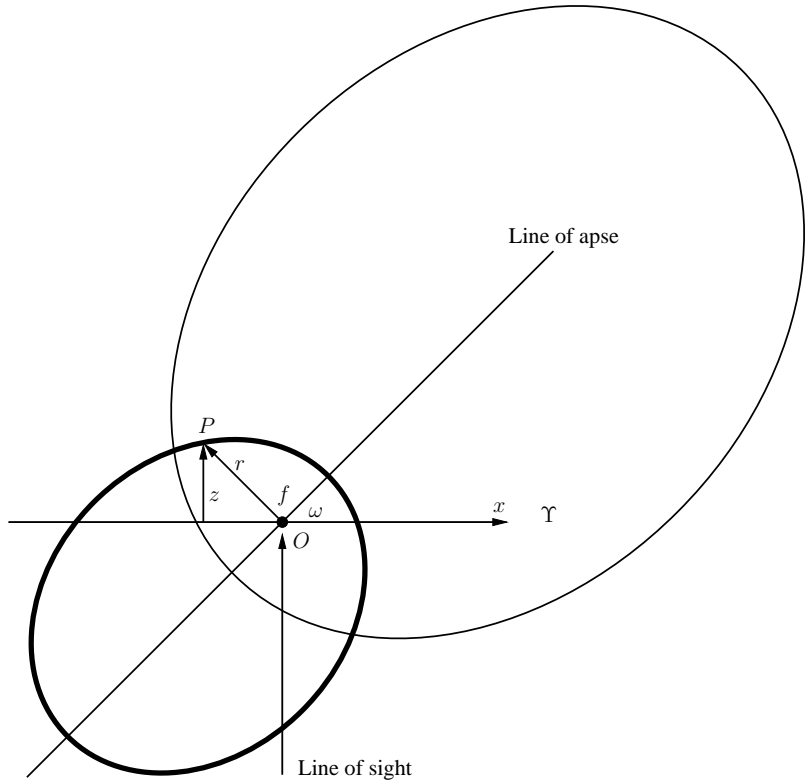}}
\caption{Graphical outline of the LITE orbit for the case $I_{b,i}=90^{\circ}$. The point $O$ is the binary-companion center of mass and denotes the origin of the coordinate system. The $x$-axis points towards the line of reference $(\Upsilon)$. The $(x,y)$-plane coincides with the plane of the sky and is perpendicular to the line of sight. The LITE binary orbit is shown as a solid ellipse with the instantaneous position of the binary at $P$. The angles $\omega$ and $f$ denote the argument of pericenter and true anomaly, respectively.}
\label{Fig0}
\end{figure*}

The effect of a single circumbinary companion on the binary period has been
presented in \citet{Irwin1952}. In that study, the eclipsing binary pair is 
regarded as a single object with a mass equal to the sum of the masses of 
each component. Because of the presence of an additional massive body (e.g. 
a stellar or planetary companion) the single binary object follows an orbit 
around the system's barycenter. This orbit, known as the light-travel time 
orbit or LITE orbit, gives rise to a modulation of the eclipsing period due to 
changes in the light-travel distance. We reproduce the LITE orbit (solid ellipse) in Fig.~\ref{Fig0} along with the orbit of the additional companion. 
\cite{Irwin1952} presents a discussion of the light-travel time orbit in a 
coordinate system with origin at the center of the LITE orbit. For reasons 
of symmetry, when discussing properties of the LITE orbit, this choice is 
suitable for the geometric interpretation of the resulting light-travel time 
curve (or $O-C$ diagram). However, a more natural choice, especially in 
systems with multiple companions, would be the system's barycenter. From Fig.~\ref{Fig0}, if $z$ measures the distance of the single binary object from the line perpendicular to the line of sight and passing through the systems 
barycenter, then the eclipsing period variation (or the observed minus 
computed, $O-C$, timing difference) will be
\begin{equation}
\tau_i = \frac{z_{i}}{c} = K_{b,i} 
\Bigg[ \frac{1-e_{b,i}^2}{1+e_{b,i}\cos f_{b,i}} \sin (f_{b,i}+\omega_{b,i}) \Bigg ],
\label{tauequation}
\end{equation}
\noindent
where $K_{b,i} = a_{b,i}\sin I_{b,i}/c$ measures the semi-amplitude of the LITE orbit in the $O-C$ diagram with $c$ denoting the speed of light and $i = 1,2$ referring to the LITE binary orbit when considering either one of the two companions. In equation (1), $a_{b,i}$ is the semi-major axis of the binary (single object) orbit, $e_{b,i}$ is its eccentricity, $I_{b,i}$ is 
the orbital inclination with respect to the plane of the sky, $\omega_{b,i}$ measures the binary's argument of pericenter and $f_{b,i}$ denotes its true anomaly. We note that we do not include any secular timing variation due to mass-transfer between the two components. As pointed out by \cite{LeeEtAl2011}, this effect is minimal and we here assume that it is negligible. Furthermore, 
it is important to note that $i=1,2$ describes two {\it separate} two-body problems. For $i=1$, we consider the binary and the inner companion whereas for $i=2$, we consider the binary and the outer companion. To obtain the resulting period variation of the measured mid-eclipse times due to both companions, one then usually assumes the superposition principle and uses the expression 
$\tau = \tau_1 + \tau_2$ for the combined light-travel time effect.

\begin{figure*}
% Tobys internal note:
% Path for "C"-Model /home/tobiash/simulations/SZHer/IDL/C-Model
% Path for "O"-Model /home/tobiash/simulations/SZHer/IDL/O-Model
\centering
\includegraphics[angle=0,scale=0.50]{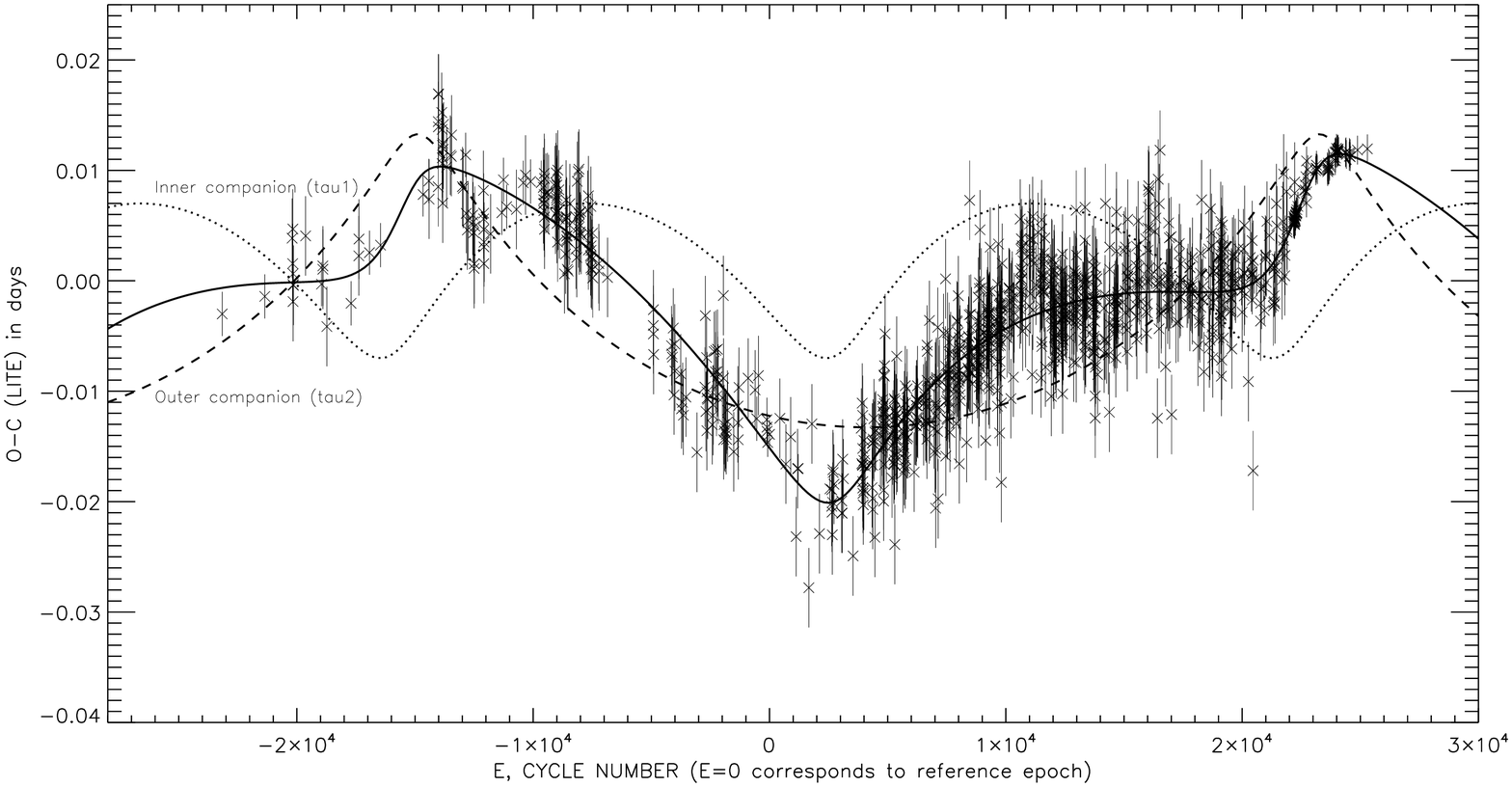}
\includegraphics[angle=0,scale=0.50]{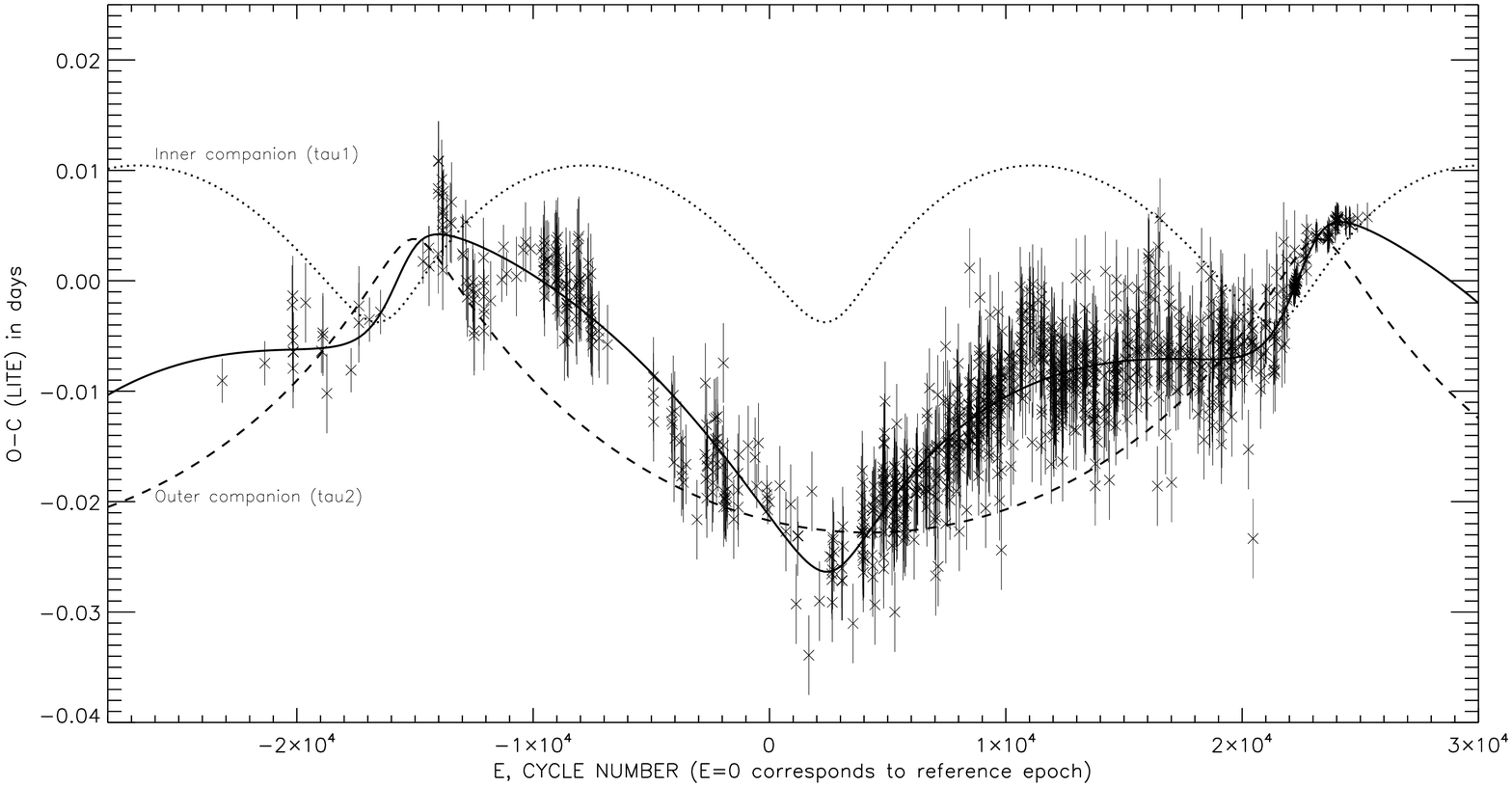}
\caption{Comparing best-fit LITE models in the ellipsoid centered (upper panel with $\chi_{r}^2 = 1.0129$) and barycentric centered (bottom panel with $\chi_{r}^2 = 1.0127$) coordinate systems. The initial guesses resulting in the two fits were taken from \cite{LeeEtAl2011}. In each panel, the inner (outer) companion is denoted by $\tau_1$ ($\tau_2$). In both panels the 
root-mean-square (RMS) scatter of data points around the best fit is around 
300 seconds.}
\label{Fig0a}
\end{figure*}

The expression in Equation (\ref{tauequation}) is obtained by evaluating the $z$-component of the binary that is along the line of sight, in a coordinate system with origin at the system's barycenter. This is different from the formulation in \cite{Irwin1952} who employs a coordinate system with origin at 
the center of the LITE orbit. The result is the omission of the $e_{b,i}\sin\omega_{b,i}$ term when comparing with the corresponding expression for the light-travel timing difference presented in \citet{Irwin1952}. We find that a barycentric coordinate system is more intuitive when carrying out the subsequent dynamical analysis. Since $\tau_{i}$ is a quantity measuring a time difference, it follows that there should be no difference in the derived orbits when formulating the light-travel timing difference in the two coordinate systems. We have tested the latter and used the best-fitting parameters from Table 6 in \cite{LeeEtAl2011} as an initial guess for the two formulations. The results are shown in Fig.~\ref{Fig0a} showing a shift along the secondary axis for the two best-fitting models. The final best-fitting orbital parameters were similar for the two models and were close to the parameters as determined by \citet{LeeEtAl2011}.

Finally, it is worth mentioning some properties of the LITE orbit for clarification. The previously mentioned Kepler elements all describe the binary system's orbit (as a single object). None of these elements are directly 
attributed to the (possibly unseen) companions orbit. The Keplerian elements of the companion(s) are inferred only indirectly from first principles. We refer 
to \cite{MurrayDermott2001} for details of properties of orbits in a barycentric coordinate system. In the following we qualitatively outline some relationships between the two orbits and refer to Fig.~\ref{Fig0}. First, the two semi-major axes are related to each other via the masses. Second, the orbital eccentricity, inclination and orbital periods $(P_{1,2})$ are the same for the single-binary object and the companions orbit. Also the apsidal lines of the two orbits are anti-aligned (see Fig.~\ref{Fig0}) giving rise to a $180^{\circ}$ difference between the two argument of pericenter angles. Orbital quantities related to a companion are denoted with a numeral subscript starting with the inner companion first (i.e $e_{1}$ shows the eccentricity of the inner companion). We note that this convention is different in \cite{LeeEtAl2011}. In their work the authors use the subscript 4 (3) to denote the inner (outer) companion.

\section{STABILITY STUDY}

\begin{figure*}
% Tobys internal note:
% Gnuplot script: plot_simpleorbit.gp
% Original EPS figure name: simpleorbits.eps
% Path: /home/tobiash/simulations/SZHer/mercury/ComputeSimpleOrbits
\centering
\includegraphics[angle=-90,scale=0.80]{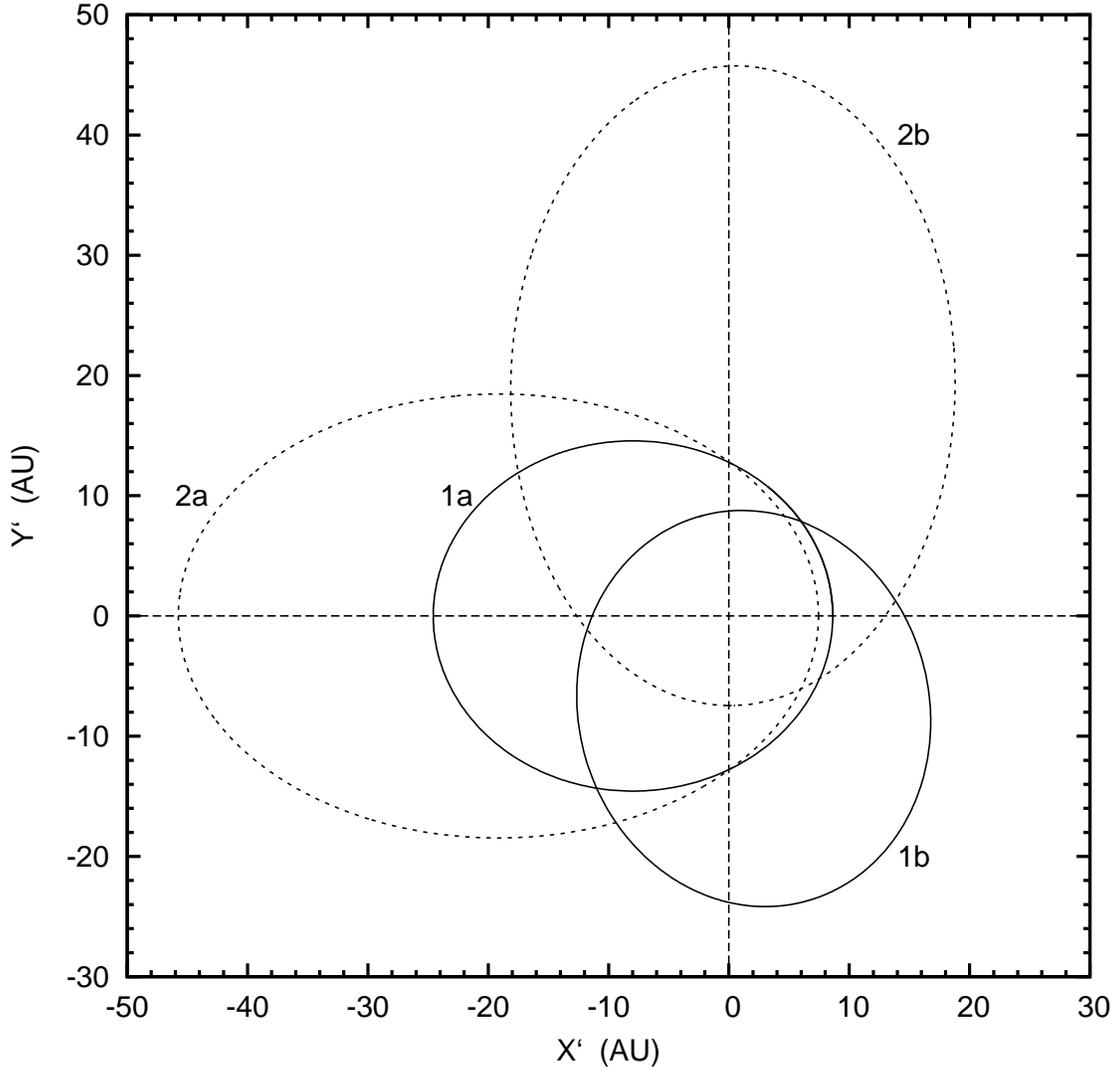}
\caption{Geometry of the two orbits (in the orbital plane) corresponding to the LITE fit parameters in Table~\ref{OrbitParams}. Both companion orbits are relative to the binarycentric reference frame (with the binary taken to be a single object) with origin at the center of the cross-hair. The orbits were integrated for one orbital period considering massless objects to visualise the two osculating single LITE orbits. We show both the $\omega = 0^{\circ}$ (1a, 2a) orbits and the orbits for $\omega = (285-180)^{\circ}$ (1b) and $\omega=(88.6-180)^{\circ}$ (2b).}
\label{Fig1}
\end{figure*}

We examined the orbital stability of the proposed three-body system (binary and
two M-type stars) of SZ Her. Orbital parameters of the two proposed companions 
are listed in Table~\ref{OrbitParams} with formal $1\sigma$ uncertainties as 
reproduced from \cite{LeeEtAl2011}. The masses and orbital semi-major axes are 
stated without formal uncertainties and were determined along with other orbital parameters, as outlined below. We show a graphical representation of the two 
osculating M-star orbits in Fig.~\ref{Fig1} for two values of the argument of 
pericenter. The figures assume the case where $I_{b,1} = I_{b,2} = 90^{\circ}$ 
with the combined binary pair placed at the origin, defining the dynamical center in an astrocentric system. An inspection of the derived osculating elements indicates that the pericenter distance $a_{1,2}(1-e_{1,2})$ 
of the inner and outer companions are at $8.63$ and $7.45$ AU respectively, implying an orbit crossing geometry. Considering the large masses of the companions, such an orbital architecture is expected to be highly unstable. 
In the following, we will study the orbital stability of the two proposed 
circumbinary companions in more detail.

We used the variable-step, Bulirsch-Stoer (BS2) $N$-body algorithm as implemented in the orbit integration package \texttt{MERCURY}\footnote{www.arm.ac.uk/$\sim$jec} \citep{MERCURY-1, MERCURY-2}. The code was compiled using the Intel Fortran Compiler\footnote{We used the following compiler flags at compile time: ifort -O3 -real-size 64} on an Linux based platform equipped with an INTEL-i5 (2.8 GHz) CPU. The initial time step was set to 0.01 days. During the integrations the maximum relative energy error was a few times $10^{-13}$.

In our calculations, we combined the masses of the two binary components and 
treated them as one single object. We used this simplifying assumption in 
order to be consistent with the orbital parameters (and minimum masses) as 
derived from the LITE formalism \citep{Irwin1952}. We integrated the orbits 
of binary and companion bodies in a ``binarycentric'' system with the combined binary mass at rest (this system is also referred to as a non-inertial astrocentric frame). This requires a transformation of orbital elements from 
the coordinate system defining the LITE orbit (barycentric frame) to the binarycentric frame. Throughout this work, we consider the two companions to 
be on the same plane. That is, $I_{1} = I_{2}$. This consideration may be justified noting that circumbinary objects may form in the same accretion disk where the binary system was formed, and as a result, orbits are more or less aligned with each other. In such a scenario, the inclination of the orbit of 
the system with respect to the plane of the sky is likely to be close to $90^{\circ}$. However, we remind the reader that no observational data exist 
for SZ Her that might allow the determination of $I_{1,2}$.

\subsection{INITIAL CONDITIONS FROM LITE ORBIT}

In general the inclination of the LITE orbit relative to the plane of the sky 
is not known. Only in the case of $I_{1,2} \simeq 90^{\circ}$ where the companion is also eclipsing one (or both) of the binary components information about this angle can be obtained from photometric measurements (e.g., Kepler-16b, \cite{DoyleEtAl2011} and Kepler-34b, Kepler-35b, \cite{WelshEtAl2012}). However, in the case of SZ Her, such a detection seems difficult due to the long LITE periods involved. Possibly third/fourth-light might be detectable from spectral data. The lack of constraints in orbital inclination implies that only information about the minimum mass $(m_{1,2}\sin I_{1,2})$ and minimum semi-major axis $(a_{1,2}\sin I_{1,2})$ can be obtained.

Information on the minimum mass from each individual LITE orbit is extracted 
somewhat similar to the case of a single-lined spectroscopic binary. The minimum mass of the two circumbinary companions were determined from a Newton-Raphson iteration using the individual massfunctions
\begin{equation}
f(m_{i}) = \frac{4 \pi^2 (a_{b,i}\sin I_{i})^3}{G P_{i}^2} = 
\frac{(m_{i}\sin I_{i})^3}{(M_{bp}+m_{i})^{2}}
\end{equation}
\noindent
for each LITE orbit and the combined mass of the binary pair $M_{bp} \approx 2.32~M_{\odot}$ \citep{LeeEtAl2011}. In agreement with \cite{LeeEtAl2011}, the masses of the two M-dwarfs were found to be $0.19~M_{\odot}$ for the inner, and $0.22~M_{\odot}$ for the outer companion (see Table \ref{OrbitParams}). 

In calculating the values of the masses, we chose not to consider the formal 
uncertainties in the companions masses since as presented by \cite{LeeEtAl2011}, these (formal) errors are on a $0.1\%$ level. From numerical experiments, these minute changes in mass have little impact on the overall dynamical evolution of the system.

The minimum semi-major axes of the LITE orbits are adopted from \cite{LeeEtAl2011} and are stated in the barycentric frame. In order to determine the semi-major axes of the two companions in the astrocentric frame, we used Kepler's third law, since the minimum mass and orbital period along 
with the total mass of the binary system are known parameters. We found the minimum semi-major axis for the inner companion to be $a_{1}\sin I_{1} = 16.6$ AU. For the outer companion we found $a_{2}\sin I_{2} = 26.6$ AU.

\subsection{RESULTS FROM ORBIT INTEGRATIONS}

\begin{figure*}
% Tobys internal note:
% Gnuplot script: plot_threefigs.gp
% Original EPS figure name: threefigs.eps
% Working directory: /home/tobiash/simulations/SZHer/mercury
\centering
\includegraphics[angle=-90, scale=0.80]{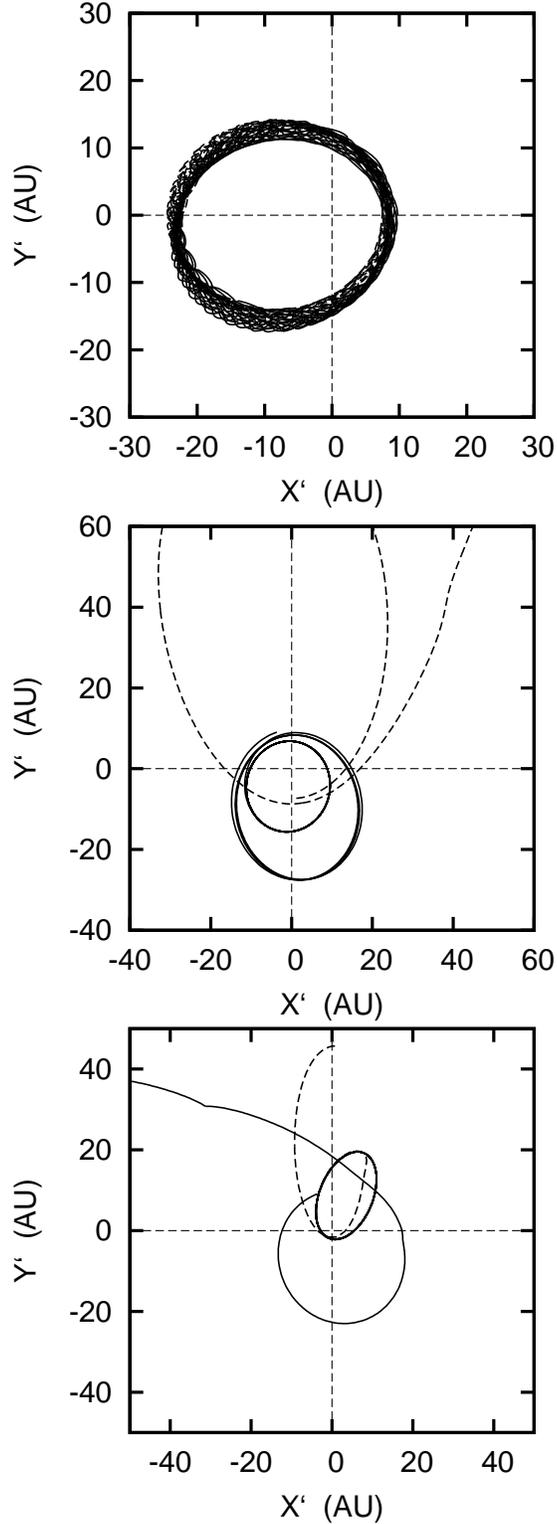}
\caption{Demonstrating unstable orbital time evolution (in the orbital plane) 
of the two companions for different initial conditions. The stipulated orbit represents the outer companion. The center of the cross-hair marks the origin of the binarycentric reference frame. In all cases we consider $I_{1,2}=90^{\circ}$. Upper panel: $\omega_{1,2} = 0^{\circ}, \lambda_{1,2} = 0^{\circ}$. Middle panel: $\omega_{1}=(286-180)^{\circ}, \omega_{2} = 
(89-180)^{\circ}, \lambda_{1,2}=0^{\circ}$. Bottom panel: $\omega_{1}=(286-180)^{\circ}, \omega_{2} = (89-180)^{\circ}, \lambda_{1}=0^{\circ}, \lambda_{2}=180^{\circ}$ }.
\label{Fig2}
\end{figure*}

In the following we present results from our stability study of the two proposed companions orbiting SZ Her. We carried out a series of short term integrations 
(up to $10^4$ years) for various initial conditions. We first considered the 
likely case of $I_{1,2} = 90^\circ$ and assigned the companions their minimum 
masses, minimum semi-major axes and eccentricites. We then varied the initial 
argument of pericenter $(\omega_{1,2})$ and initial mean longitude 
$(\lambda_{1,2})$ for the two companion M-type stars and integrated their 
orbits. Figure \ref{Fig2} shows the results of a few integrations. The upper 
panel in this figure shows the case where the orbital apsidal lines of the two 
companions are initially parallel and both companions start with the same mean 
longitude ($\lambda_{3,4}=0^{\circ}$). The result is a quick capture of the two 
companions into a binary system. Although this initial condition results in a 
long-term stability, it does not reflect the derived Keplerian parameters of the 
two LITE orbits proposed by \cite{LeeEtAl2011}. The middle panel of
Fig.~\ref{Fig2} shows the result of an integration where the outer circumbinary 
component escapes the system. In this integration, we used the fitted argument 
of pericenter for each LITE orbit to derive the argument of pericenter for the 
two companions. Finally, we considered the case where the outer M-dwarf companion is started at its apocenter $a_{2}(1-e_{2})$ with $\lambda_{2} = 180^{\circ}$. The remaining Kepler elements were chosen similar to the previous case. The corresponding orbits are shown in the lower panel of Fig.~\ref{Fig2}. As shown here, for this set of initial conditions, the inner companion escapes from the system in less than 100 years.

In general, for the escape scenarios, because of the conservation of energy, 
one companion is transfered to a smaller orbit as a result of the escape of the other companion. We encountered this behaviour in all our integrations where 
an escape took place. We then considered various orbits with semi-major axes and eccentricities within and beyond their formal $1\sigma$ uncertainties as given by \cite{LeeEtAl2011}. The majority of these integrations (especially those with higher orbital eccentricities) resulted in unstable orbits. Low-eccentricity orbits showed a somewhat longer lifetime. The main mechanism resulting in the system's instability was the escape of one of the companions as a result of a 
close encounter with the other.

\begin{figure*}
% Tobys internal note:
% Gnuplot script: plot_sixfigs.gp
% Original EPS figure name: sixfigs.eps
% Working directory: /home/tobiash/simulations/SZHer/mercury
\centering
\includegraphics[angle=-90, scale=0.52]{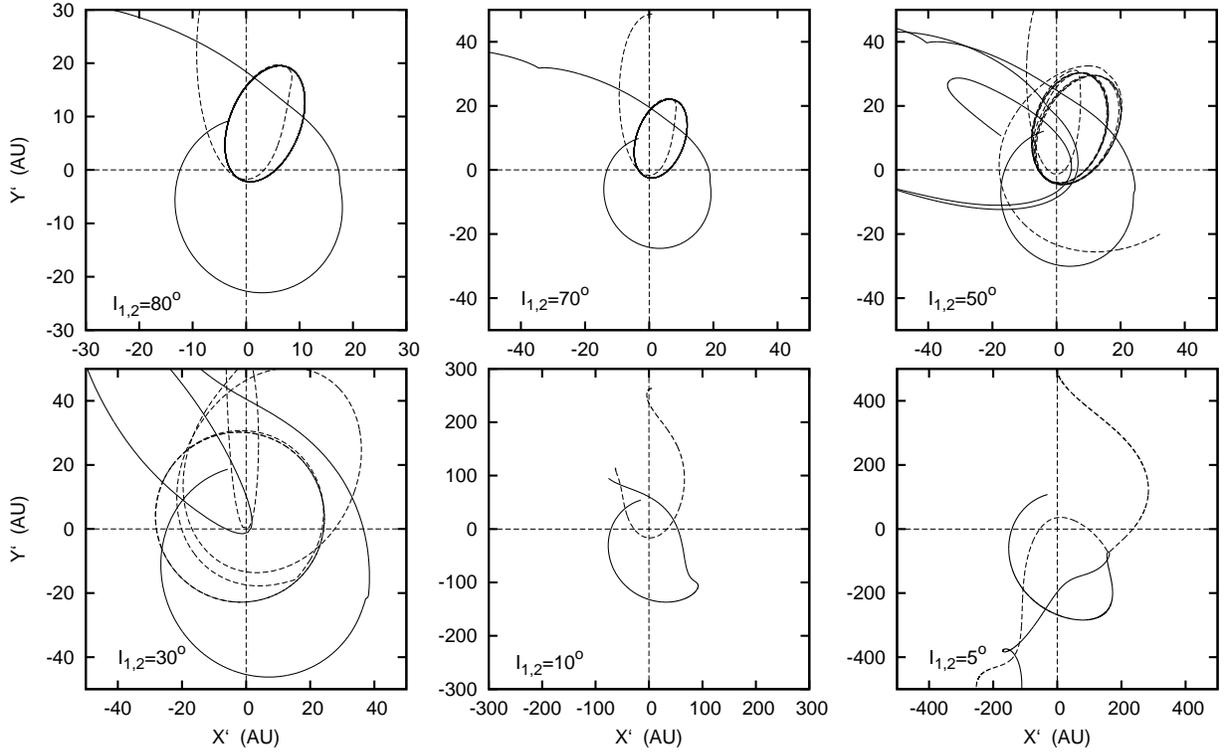}
\caption{Unstable orbits from numerical integrations for various LITE orbital 
inclinations. Each panel shows the orbits (in the orbital plane) using initial conditions that considers a scaled semi-major axis and mass for the two companions simultaneously. The inclination between the line of sight and the plane of the sky were $I_{1,2} = 5^{\circ},10^{\circ},30^{\circ},50^{\circ},70^{\circ},80^{\circ}$. 
In all integrations, $\lambda_{2}=180^{\circ}$. The orbits of the two companion were assumed to be co-planar. The stipulated line always represents the outer binary companion. The centre of the hair-cross marks the origin of the barycentric reference frame.}
\label{Fig3}
\end{figure*}

In order to search for a stable orbital configuration which reflects the main 
characteristic of the two proposed one-companion LITE orbits, we considered initial conditions with different LITE orbit inclinations for the two companions on co-planar orbits. Although the orbital separations increases with decreasing inclination, mutual gravitational perturbations are still important since the masses of the two companions increase as well. Therefore, it is not obvious apriori if a stable configuration can be found. We integrated the equations of motion for ten different values of the inclination with respect to the plane of the sky in the range of $I_{1,2} \in [1^{\circ}, 90^{\circ}]$. For each inclination, we scaled the minimum mass and semi-major axis for the two companions accordingly in order to obtain their true (assumed) mass. In each integration, we placed the outer component at $\lambda_{2} = 180^{\circ}$. A representative subsample of our results is shown in Fig.~\ref{Fig3}. In all ten cases we found the system to be highly unstable on short timescales. To double-check our results, we repeated the numerical integrations using the RADAU algorithm (also available in the \texttt{MERCURY} package). The outcome confirmed our previous results. 

In each integration, we found multiple close approaches that either led to ejections or collisions between the two proposed circumbinary companions. We repeated our integrations with various initial $\lambda_{1,2}$ resulting in no difference in the overall inferred orbital instability. Considering a wide range of orbital possibilities, we find that all our orbital integrations to result in highly unstable systems.

\section{METHODS AND RESULTS FROM REVISED LITE MODELS}

In order to find a possibly stable LITE model to the proposed quadruple system, we carried out an extensive search of the $\chi^2$ parameter space. The  analysis, methodology and technique are similar to those described in detail in \citet{HinseEtAl2011}. The only exception is that we now formulate the LITE model in the barycentric frame by omitting the $e\sin\omega$ term in equation 3 of \citet{HinseEtAl2011} as outlined earlier. In the following, we briefly repeat main aspects of the underlying analysis.

We used the Levenberg-Marquardt least-square minimisation algorithm as  implemented in MPFIT \citep{Markwardt2009} software. The goodness-of-fit  statistic of each fit was evaluated from the weighted sum of squared errors,  $\chi^2$. Here we use the reduced chi-square statistic $\chi_{r}^2$. We seeded 107,625 initial guesses within the framework of a Monte Carlo experiment.

Each guess was allowed a maximum of 500 iterations before termination. Converged solutions were accepted with initial guess and final fitting parameters recorded. Initial guesses of the model parameters were chosen at random from either a uniform or Gaussian distribution. \citet{LeeEtAl2011}, for example, provided a Lomb-Scargle period analysis on the complete timing data set. They determined two possible dominant frequencies of 
$3.53 \times 10^{-5}~\textnormal{cycle}~d^{-1}$ and 
$7.06 \times 10^{-5}~\textnormal{cycle}~d^{-1}$ associated with the two proposed circumbinary companions. The shorter period is relatively well determined given the long observational time span of SZ Her. We, therefore, draw random periods from a Gaussian distribution with the standard deviation three times larger for the longer period. The standard deviation for the shorter period was set to $\pm 8$ years with its mean centered at 43 years. This choice somewhat restricts the quasi-global exploration of the (in principle infinite) $\chi^2$-parameter 
space to within a physically meaningful sub-region.

We used the same timing data set as in \cite{LeeEtAl2011}, but applied our model on times defined by a uniform time standard. In \cite{LeeEtAl2011}, timing data for SZ Her were recorded in the UTC time frame, which is known to be non-uniform \citep{GuinanRibas2001}. We therefore transformed the HJD (Heliocentric Julian Date) timing records in UTC (Coordinated Universal Time) into the terrestrial time (TT) standard \citep{Bastian2000}. The resulting unit of TT time is HJED (Helocentric Julian Ephemeris Date).

\subsection{Results}

The results from our Monte Carlo experiment are somewhat similar to the results 
presented in \cite{HinseEtAl2011}. We discarded all guesses which reached a lower or upper boundary in one of the model parameters (since no formal errors are supplied for these parameters within the MPFIT algorithm). The total number of qualified guesses were then reduced to 30,700. The majority of guesses (15,659 or 50\%) had final reduced goodness-of-fit parameter in the interval $1.0086 \le \chi_r^{2} < 1.0186$. This is in agreement with the two-LITE model found in \cite{LeeEtAl2011}. However, a small number (3758 or 12\%) of initial guesses had a final reduced goodness-of-fit statistic in the interval $0.9886 \le \chi_r^{2} < 0.9986$ with the best-fitting model resulting in $\chi_{r}^2 = 0.9886$. All remaining fits had $\chi_r^{2} > 1.05$.

\begin{figure*}
% Tobys internal note:
% Working directory: /home/tobiash/simulations/SZHer/IDL/SinglePlotBestFits
% IDL script: ltt_cycle.pro
% Original EPS file name: BestFit_Chi2_0.989.eps
\centering
\includegraphics[angle=0, scale=0.50]{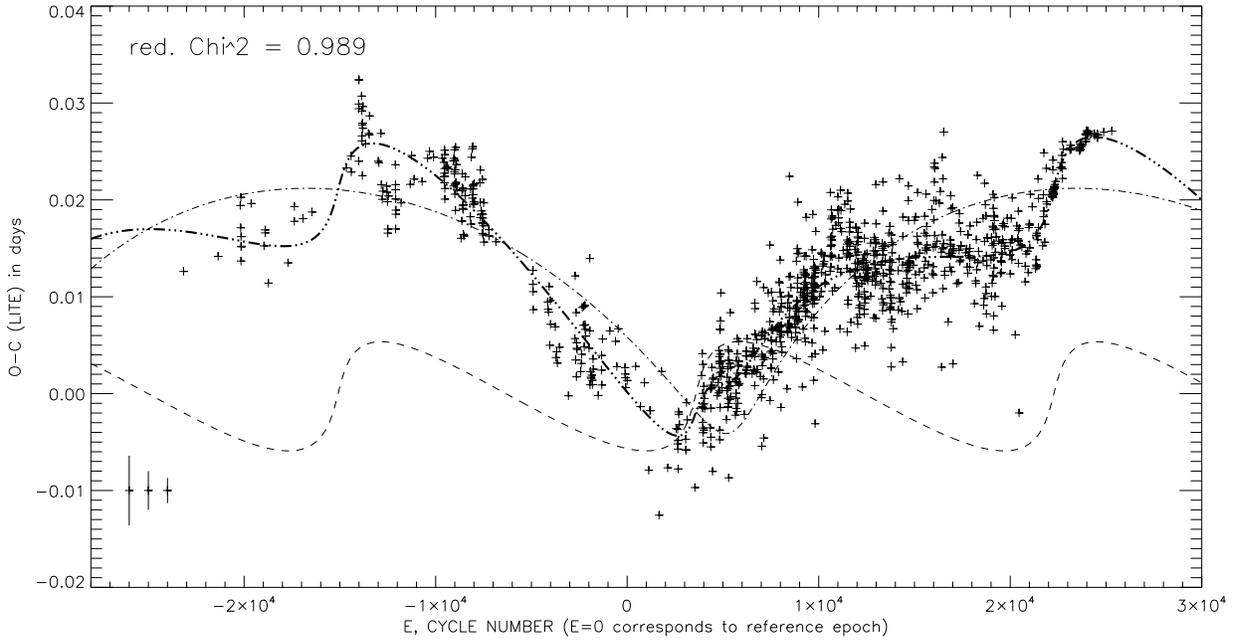}
\caption{Best fit model with $\chi_{r}^2 = 0.989$ (dash-dot-dot-dot) from our 
many-guess Monte Carlo experiment and the two companion sinusoidal-like variations: inner (dash) and outer companion (dash-dot). The corresponding orbital parameters for the two LITE orbits are shown in Table \ref{bestfitparam}. The root-mean-square scatter of data around the best fit is $\sim 292$ seconds. We show the three $\pm 1\sigma$ timing error bars in the lower left corner corresponding to 0.0036, 0.0020 and 0.0013 days as adopted by \cite{LeeEtAl2011} for various observation techniques.}
\label{Fig4}
\end{figure*}

Our best-fitting model is somewhat smaller than the two-LITE fit presented in 
\cite{LeeEtAl2011}. We show the best-fitting model in Fig.~\ref{Fig4} with the 
best fitting parameters listed in Table \ref{bestfitparam}. We note that the 
orbital eccentricity of the inner companion increased significantly from 0.48 to 0.76 when comparing with the work in \cite{LeeEtAl2011}. Orbital radii, periods, as well as the minimum masses of the two companions are almost unchanged. We calculated the mean value and corresponding standard deviation of the final semi-major axis and eccentricity for the two LITE orbits. For the inner LITE orbit, the average final semi-major axis and eccentricity were $1.24 \pm 0.38$ AU and $0.57 \pm 0.17$, respectively. For the outer LITE orbit, the average final semi-major axis and eccentricity were $2.32 \pm 0.41$ AU and $0.59 \pm 0.22$.

\subsection{ORBIT STABILITY OF BEST-FIT AND LITE COMPUTATION}

\begin{figure*}
%Figure info:
%Path:/home/tobiash/simulations/SZHer/mercury/BESTFITORBITSTABILITY
%Xfig file: fourfig_xfig.fig
%Within Xfig: Use special flag and export to EPS/LaTeX.
%Note: the line \usepackage[dvips]{color} is necessary to make this work.
\centering
\scalebox{0.75}{\input{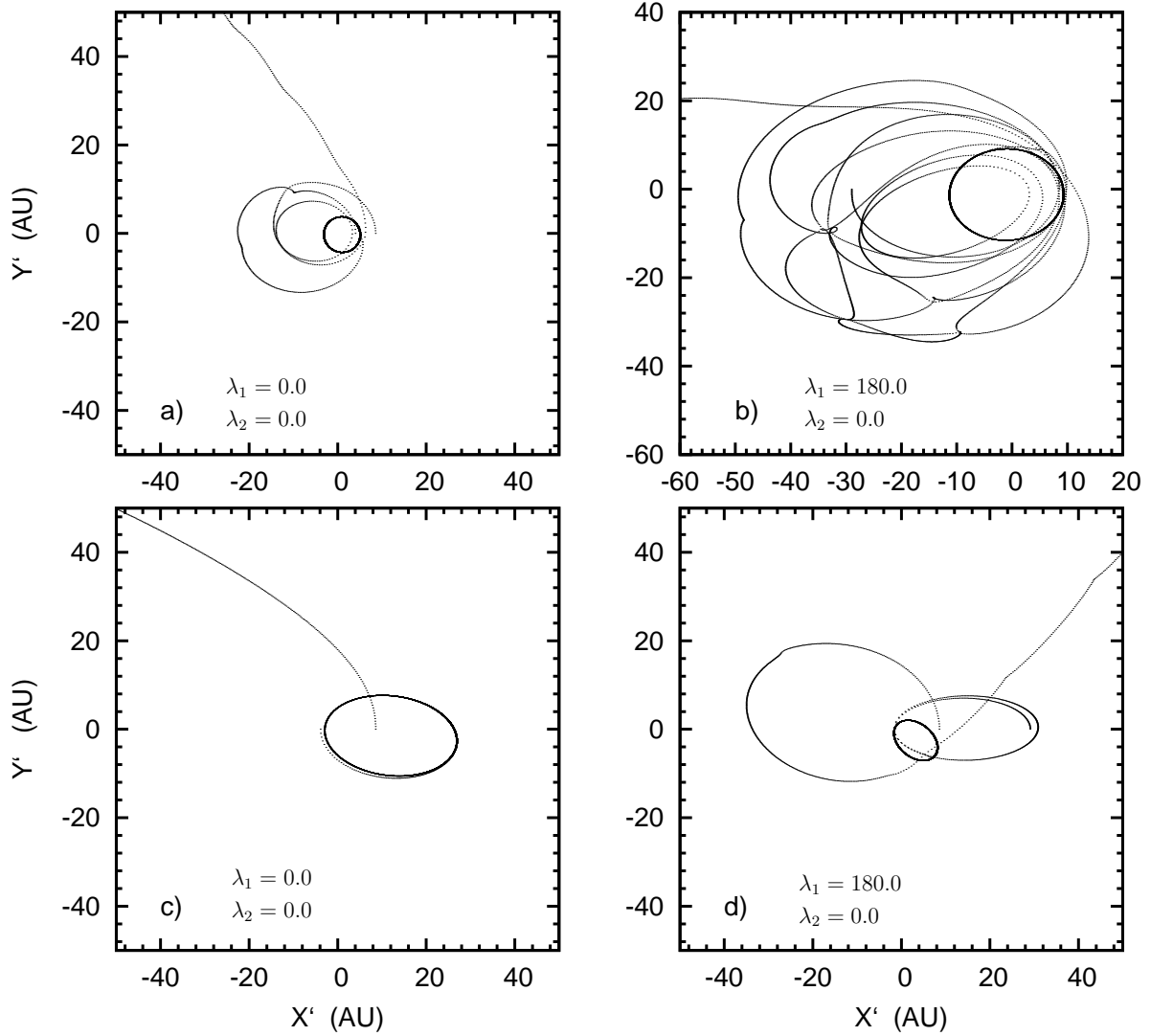}}
\caption{Time evolution of the two M-dwarf companions using initial conditions from Table \ref{bestfitparam}. The mean longitude is denoted by $\lambda$. Upper two panels are for $\omega_{1}=0$ and bottom two panels for $\omega_{1} = 180^{\circ}$. In each panel the binary pair is placed at the origin of the coordinate system.}
\label{Fig5}
\end{figure*}

In the previous section, we obtained an improved fit to the existing timing 
data of the eclipsing SZ Her system. The resulting best-fit osculating orbits of the two proposed companions showed an orbit-crossing architecture. To examine the orbital stability of the best-fit quadruple system the equations of motion were integrated in an astrocentric system using \texttt{MERCURY}. The binary pair was treated as a single object. We considered a large combination of different initial conditions and studied the final outcome for different values of semi-major axis, eccentricity and orbital angular variables. In particular, we studied the dynamics of the system near the suggested 2:1 mean-motion resonance. All integrations were carried out for 10000 years. Total system energy was conserved to within a few times $10^{-12}$. In all cases, integrations resulted in the escape of one of the two companions from the system. We show four examples of the time evolution of the orbits of the two M-dwarfs in Fig.~\ref{Fig5}. It was assumed, initially, that the two companions were co-planar ($I_{1,2} = 90^{\circ}$) with the binary system. The upper (lower) two panels of Fig.~\ref{Fig5} show the case for $\omega_1 = 0$ 
($\omega_1 = 180$ degrees). In Fig.~\ref{Fig5}c, the outer companion escapes the system within a few years. The resulting system consists of a single companion on a stable Keplerian orbit causing (at most) a single sinusoidal LITE effect.

\begin{figure*}
% Figure info:
% Path:/home/tobiash/simulations/SZHer/mercury/COMPUTETWOLITESIGNAL
% Gnuplot script: plot_fourfigs_ltt.gp
% Original EPS figure name: fourfigs_ltt.eps
\centering
\includegraphics[angle=-90, scale=0.57]{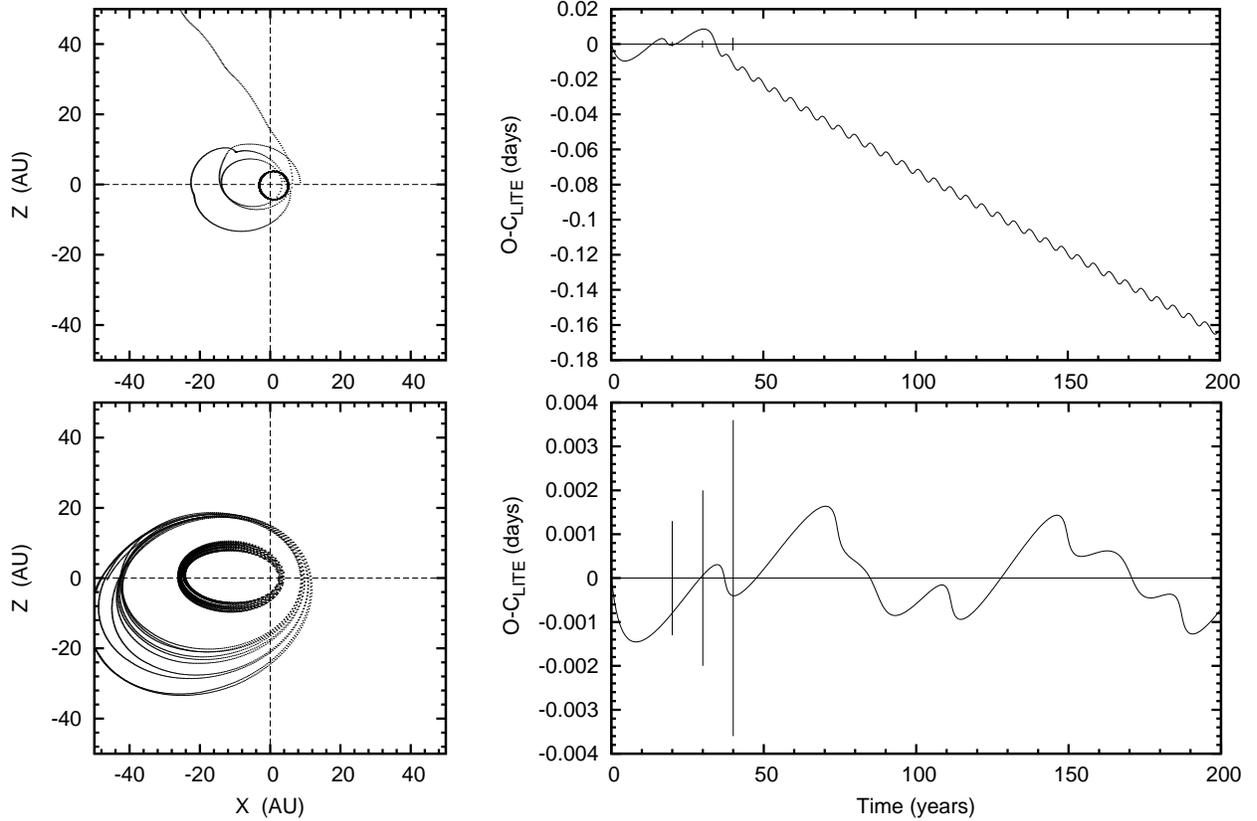}
\caption{Numerical computation of the orbit (left) and the resulting LITE effect (right) for two different scenarios of companion mass. Initial conditions are the same as in Fig.~\ref{Fig5}a with $I_{1,2} = 90^{\circ}$. Upper panel: Companions masses are $m_1 = 0.23~M_{\odot}$ and $m_{2} = 0.21~M_{\odot}$. Lower panel: Companions with masses of $m_1 = 0.023~M_{\odot}$ and $m_{2} = 0.021~M_{\odot}$. Vertical bars in the right panels represent $\pm 1\sigma$ uncertainties with $\sigma$ corresponding to 0.0013 (112 seconds), 0.0020 (173 seconds) and 0.0036 days (311 seconds).}
\label{Fig6}
\end{figure*}

To demonstrate this single sinusoidal variation, we computed the light-travel 
time effect from a direct numerical integration for two smaller values of the two companions masses. We considered the case for which the two companions are co-planar with the binary plane ($I_{1,2} = 90^{\circ}$). The upper panel of Fig.~\ref{Fig6} shows an unstable system and is similar to the orbit shown in Fig.~\ref{Fig5}a. As explained earlier, one companion is ejected from the system within 50 years. The resulting LITE effect exhibits an initial variation in the binary period. Because of the conservation of the total linear momentum, the binary (single object) and bound companion move in a direction opposite to that of the ejected companion. The result is a one-component LITE effect superimposed on a constant period change (linear trend) due to the systemic motion of the 
whole system towards the observer (negative $O-C$ values). The systemic velocity is obtained from the slope of the linear trend. The lower panel of Fig.~\ref{Fig6} shows the orbits of the two companions with their mass reduced by a factor of ten. The initial conditions in these simulations are similar to those in the upper panel of Fig.~\ref{Fig6}. Because of the smaller mass of the circumbinary companions, the system is now stable on a 200 year time scale. However, for longer integration times, the outer companion is ejected from the system as a result of strong mutual interactions with the inner one. 

Within the 200 year integration the resulting LITE effect shows a quasi-periodic modulation of the binary period. The semi-amplitude $K_{b,i}$ of this signal is around 0.0015 days or 130 seconds. In order to reliably detect such a 
LITE signal would require $1\sigma$ timing uncertainties to be significantly smaller than 112 seconds (0.0013 days).

\section{CONCLUSION \& DISCUSSION}

In this work, we have reanalysed the complete timing data set of SZ Her and 
determined an improved two-LITE fit using an extensive Monte Carlo based 
$\chi_{r}^2$-parameter space search. Our model used the center of mass of 
the one-companion system as the origin of the underlying coordinate system. 
Using this new approach, we find no significant change in the derived 
parameters when comparing our results with those presented in \cite{LeeEtAl2011}. As shown in this work, the LITE orbits derived from the 
two coordinate systems (one with origin in the center of the LITE orbit 
which includes $e\sin\omega$ term and the other with origin in the  system's barycenter as expressed by Equation (\ref{tauequation})) are similar and no significant difference was observed between the two models.

In comparison with \cite{LeeEtAl2011} and assuming the two-LITE model is 
correct, our fit with $\chi_{r}^2 = 0.989$ provides a slightly better description of the underlying timing data. The existence of the improved fit presented in this work is most likely explained by our thorough (quasi-global) $\chi_{r}^2$ parameter search. However, the majority of initial guesses resulted in a slightly larger $\chi_{r}^2$ statistic, which is consistent with the first fit found in \cite{LeeEtAl2011}. To some extent, the fit found by \cite{LeeEtAl2011} is more ``stability-friendly'' due to the lower eccentricity of the inner proposed companion. Furthermore, our fit also 
suggests the two proposed companions to be in a near 2:1 mean-motion resonance 
with periods of 42 and 90 years for the inner and outer orbits. However, we have considered various initial configurations close to this resonance all of which resulted in unstable orbits.

Our stability analysis showed that all models presented in \cite{LeeEtAl2011} 
and in this work resulted in unstable quadruple systems (considering the three-body problem). In particular, our parameter study of the unknown orbital inclination with respect to the line of sight indicated that reducing the inclination from $90^{\circ}$ to $1^{\circ}$ results in an increase in the companions semi-major axis. However, at the same time, the companion's masses also increas which in turn introduces large gravitational perturbations between these two objects. Our inclination survey concluded that all low-inclination orbits also result in unstable systems on very short time scales. 

If the observed timing variation is real and caused by the presence of two 
circumbinary objects, then the system needs to be in a stable configuration. 
We therefore conclude that either the two-LITE model is an inadequate 
description to the data, and/or the underlying data set is insufficient. For 
that reason, in order to constrain any future modeling efforts, we encourage 
the aquisition of additional photometric observations of SZ Her. 

There may also be an alternative, although unlikely, explanation for the observed period modulation in the SZ Her binary orbit as originally outlined by \cite{HornerEtAl2011}. This possibility depicts a scenario for which the two 
companions currently undergo a dramatic dynamical evolution with a transition 
from a stable to an unstable configuration with one companion escaping the 
system shortly as suggested in this work. The dynamical reason for this instability would remain an open question and additionally renders this scenario unlikely. If true, the observational consequence of such a scenario should make it possible in the foreseeable future, to detect a linear trend in the measured 
$O-C$ diagram. This was demonstrated numerically in the top panel of 
Fig.~\ref{Fig6}. However, due to the short orbital instability time scales 
found in this work, it seems unlikely that we are currently witnessing a break-up scenario in which the system enters the very endstate of its dynamical 
evolution with one companion on the verge of escaping the binary system. The 
leading question is: why should we be in the fortunate situation and observe 
the last few hundred years of a supposedly long-lived (possibly millions of 
years) system that gradually evolved towards a general instability? 

On the other hand, and as mentioned earlier, a dynamical inspection of 
currently known proposed circumbinary planets reveals these systems are 
unstable as well (at least three out of four). This finding could be attributed to the fact that ground-based photometric observations are more sensitive to detect sub-stellar circumbinary objects introducing large-amplitude period variations superimposed on the linear ephemeris of the mid-eclispe times of the binary. In this case the quintessence would be: larger masses would introduce larger perturbations and cause larger period modulations in the $O-C$ diagram and therefore such systems would be prone to disintegrate on a short time-scale due to strong mutual interactions. Another argument for the non-existence of the two components with orbits proposed in \cite{LeeEtAl2011} comes from numerical $N$-body orbit calculations. Studies of the dynamical stability of hierarchical four-body systems were carried out by \cite{SzellStevesErdi2004}. Their work suggests that low-mass stellar objects on circumbinary orbits result in unstable hierarchical systems. Considering symmetric pairs of masses, these authors showed that for large mass-ratios, the most likely event is a double binary configuration (e.g. HD98800, \cite{FurlanEtAl2007}). Indeed, in one of our experiments, we confirmed such an outcome from a direct numerical integration. On the other hand, for small mass-ratios (possibly comparable to planetary masses), their work suggest that the most likely outcome are circumbinary orbits. This again was demonstrated when decreasing the masses of the two companions by a factor of 10. However, a thorough stability analysis of circumbinary four-body low-mass stellar systems, considering a large range of orbital parameters and masses, would be helpful to identify stable domains of circumbinary orbits.

It isimportant to note that other sources may exist that can create modulations in the observed binary period. This could be in the form of magnetic interaction between the two binary components, and/or mass or angular momentum transfer resulting in a secular modulation of observed timings. In their study, \citet{LeeEtAl2011} point out the possibility that SZ Her (being a semi-detached binary pair with the less massive component filling its Roche lobe) is currently undergoing a phase of weak mass transfer. However these authors provide arguments that this effect is likely to be negligible. Other mechanism potentially capable of causing eclipse timing variations is the direct gravitational perturbations by a companion on the binary orbit. This 
possibility was not considered in \cite{LeeEtAl2011} and is left for a future study.

Star-spots also could introduce stellar jitter mimicing period variations as discussed in \citet{WatsonDhillon2004}. Also, timing measurements might 
suffer from systematic measurement errors introducing correlated red noise possibly resulting in wrong model parameters \citep{ColesEtAl2011}. Unaccounted systematic (red) timing errors resulted previously in the false 
detection of planets around pulsars.

To unveil the true nature of the observed timing variation we encourage 
the aquisition of future photometric/spectroscopic follow-up observations 
of SZ Her allowing to further constrain and refine timing models. Future models 
favouring a two-companion solution should be tested for orbital stability 
and the resulting $O-C$ variation obtained from numerical integrations 
compared with the inferred timing measurements.

\subsubsection*{Acknowledgements}

Research by TCH is carried out at the Korea Astronomy and Space Science 
Institute (KASI) under the KRCF (Korea Research Council of Fundamental Science 
and Technology) Young Scientist Research Fellowship Program. Numerical
simulations were carried out on the ``Beehive'' Computing Cluster at Armagh 
Observatory (UK). TCH acknowledges Martin Murphy for 
assistance in using the Beehive computing cluster and Professor Chun-Hwey 
Kim (Chungbuk National University, Cheongju, South Korea) for stimulating 
discussions on eclipsing binaries and their period variations. KG is supported by the Polish Ministry of Science and Higher Education through grant N/N203/402739. NH acknowledges support from NASA Astrobiology Institute under Cooperative Agreement NNA04CC08A at the Institute for Astronomy, University of Hawaii, and NASA EXOB grant NNX09AN05G. JWL and CUL acknowledges support from KASI registered by grant number 2012-1-410-02. Astronomical research at Armagh Observatory is funded by the Department of Culture, Arts and Leisure (DCAL).

\clearpage

\begin{table}
\centering
\begin{tabular}{lccc}
\hline
Parameter& SZ Her(AB)C & SZ Her(AB)D &Unit \\
\hline
\hline
minimum semi-major axis, $a_{1,2}\sin I_{1,2}$ & 16.6 & 26.6 & AU \\
eccentricity, $e_{1,2}$ & $0.48 \pm 0.17$ & $0.72\pm 0.09$ & - \\ 
argument of pericenter, $\omega_{1,2}$ & $105 \pm 10$ & $268.6 \pm 7.5$ & degrees \\
orbital period, $P_{1,2}$ & $42.5\pm 1.1$ & $85.8\pm 1.0$ & year \\
minimum mass, $m_{1,2}\sin I_{1,2}$ & $0.188$ & $0.222$ & $M_{\odot}$ \\
\hline
\end{tabular}
\caption{Binarycentric or astrometric orbital parameters of the two circumbinary companions C and D. Note that we have accounted for the $180^{\circ}$ difference in the argument of pericenter angle. Parameters with formal 1$\sigma$
uncertainties are from \cite{LeeEtAl2011}. All other parameters are obtained as outlined in the text.}
\label{OrbitParams}
\end{table}

\clearpage

\begin{table*}
% Toby's internal note:
% amplitude, massfunction, minimum mass, and projected semi-major axis and 
% associated erros have been calculated using the IDL script: minimass.pro
% Path on kapc: /home/tobiash/simulations/LTT/idl
\centering
\begin{tabular}{lcccc}
\hline
Parameter                && \multicolumn{2}{c}{two-LITE}                   &   Unit            \\ [1.5mm] \cline{3-4} \\ [-2.0ex]
                         && $\tau_{1}~(i=1)$                 & $\tau_{2}~(i=2)$            &                   \\ 

\hline
$\chi_{r}^2$            &&  \multicolumn{2}{c}{0.989}                          &   -               \\
\hline
RMS            &&  \multicolumn{2}{c}{0.00339}                          &   days               \\
\hline
$T_0$                    &&  \multicolumn{2}{c}{2,434,987.38455(31)}           &   HJED             \\
$P_0$                    &&  \multicolumn{2}{c}{0.818095801(11)}              &   days               \\
$a_{b,i}\sin I_{b,i}$      &&  $1.49\pm 0.53$                 &  $2.23 \pm 0.41$            &   AU              \\
$e_{b,i}~~(\textnormal{or}~e_{1,2})$                      &&  $0.76 \pm 0.14$                   &  $0.69 \pm 0.12$              &   -                \\
$\omega_{b,i}$                 &&  $2.42 \pm 7.1$                    &  $286.16 \pm 10.1$               &   deg             \\
$T_{b,i}$                      &&  2,422,649(322)              &    2,439,551(198)         &   HJED             \\
$P_{b,i}~~(\textnormal{or}~P_{1,2})$                &&  $15286 \pm 224$              &  $32860 \pm 341$              &   days            \\
\hline
$K_{b,i}$                      && $0.00864(32)$ & $0.00129(22)$ &   days                            \\
$f(m_{i})$ && $0.00191(65)$ & $0.00138(37)$ & $M_{\odot}$ \\
\hline
$m_{i} \sin I_{i}$       && $0.23 \pm 0.08$                 & $0.21 \pm 0.05$            &  $M_{\odot}$       \\
$a_{i}\sin I_{i}$          && $16.5(9)$                     & $27.4(8)$                &  AU                       \\
$e_{i}$                 && $0.76 \pm 0.14$ & $0.69 \pm 0.12$ & - \\
$\omega_{i}$ && $182 \pm 7.1$ & $106 \pm 10.1$ & deg \\
$P_{i}$ && $15286 \pm 224$ & $32860 \pm 341$ & days \\
\hline
\end{tabular}
\caption{Best-fit parameters for the LITE orbits of SZ Her corresponding 
to Fig.~\ref{Fig4}. Subscripts $1,2$ 
refer to the circumbinary companions with $i=1$, the inner, and $i=2$, the outer, companions. RMS measures the root-mean-square scatter of the data around the best fit. Formal uncertainties obtained from the covariance matrix are valid for the last digit and shown in paranthesis. Note that the eccentricity and orbital period are shared quantities as outlined in the text. The last five lines are quantities of the two companions in the astrocentric coordinate system.}
\label{bestfitparam}
\end{table*}

\clearpage

\clearpage

\clearpage

\clearpage

\clearpage

\clearpage

\clearpage

\clearpage

\end{document}